\documentclass[aps,titlepage,preprint,12pt,nofootinbib]{revtex4-2}

\usepackage{amssymb}
\usepackage{amsfonts}
\usepackage{setspace} 
\usepackage{graphicx} 
\usepackage{subfigure}
\usepackage[margin=1in]{geometry}
\usepackage{amsthm}
\usepackage{mathrsfs} 
\usepackage{caption}
\usepackage{mathtools}
\usepackage{amsmath}
\usepackage{hyperref} 
\hypersetup{
    colorlinks=true,
    linkcolor=blue,
    filecolor=blue,      
    urlcolor=blue,
    citecolor=blue,
    }
\begin{document}

\title{On the impossibility of observational confirmation of black holes} 

\author{Thiago T. Bergamaschi}
\email{tbergamaschi@ifsc.usp.br}
\affiliation{Sao Carlos Institute of Physics, University of Sao Paulo, IFSC – USP,
13566-590, Sao Carlos, SP, Brazil}
\date{\today}

\begin{abstract}
General relativity has achieved remarkable experimental and observational success. Critically, recent data from the LIGO-Virgo-KAGRA, Event Horizon Telescope, and GRAVITY collaborations are often credited with \textit{demonstrating} the existence of black holes, but in fact they only provide evidence for objects that should be regarded as black hole candidates. While current data are in striking agreement with the predictions for Kerr black holes, they can only rule out specific alternative models of compact objects rather than provide conclusive proof of black holes. More fundamentally, and independent of whether or not black holes exist, general relativity itself imposes limits on what can be observationally established. Essentially, no observational data is sufficient to confirm the existence of black holes.

   \vspace{24pt}
\noindent Essay written for the Gravity Research Foundation 2026 Awards for Essays on Gravitation.

\end{abstract} 

\maketitle
%%%%%%%%%%%%%%%%%%%%%%%%%%%%%%%%%%%%%%%%%%%%%%%%%%%%%%%%%%%%%%%%%%%%%%%%%%%%%%%%%%%%%%%%%%%%%%%%%%%%%%%%%%%%%%%%%%%%%%%%%%%%%%%%%%%

\subsubsection*{Introduction}

In general relativity, black holes are defined as causally disconnected regions, i.e., from which no information can escape to infinitely distant regions. Although this definition is inherently teleological, as it relies on global properties of spacetime \cite{Wald1984}, numerical simulations together with the uniqueness theorems \cite{Heusler1996, Chruściel2012} suggest that physical black holes (e.g., those resulting from the gravitational collapse of a star or compact-object mergers) should be described by the Kerr metric once they reach a stationary configuration. Essentially, observers outside a black hole can obtain information only about physics outside of it, which should consist of a perturbed Kerr spacetime due to gravitational-wave emission and nearby small matter-energy sources. While some have argued that this mathematical definition is unphysical in astrophysical contexts \cite{Frolov2014, Thorne1986}, black holes have nonetheless catalyzed major advancements in theoretical physics over the past fifty years, serving as a conceptual laboratory for the interplay of gravitation, thermodynamics, and quantum theory (see, e.g., \cite{Wald2001, Parker2009, Bergamaschi2026}).

Over the past decade, black holes have entered the observational spotlight, largely due to the efforts of the LIGO-Virgo-KAGRA (see, e.g., \cite{Abbott2016, Abbott2018, Abac2025}), Event Horizon Telescope (see, e.g., \cite{EHT2019}), and GRAVITY collaborations (see, e.g., \cite{GRAVITY2018}), which have analyzed gravitational and electromagnetic data with unprecedented precision. These observations have opened extraordinary prospects for testing fundamental physics, but it is imperative to understand what such data can probe, falsify, and \textit{confirm}.

This distinction is particularly salient for black holes because, while general relativity imposes no theoretical limit on how close one may approach a black hole's boundary (i.e., the event horizon) and still communicate with the rest of the universe, practical observational thresholds quickly become relevant. The central challenge lies in distinguishing a black hole from an ultracompact object. When an ultracompact object is characterized by a sufficiently small closeness (or cutoff) parameter\footnote{This parameter is a measure of how ``close'' a compact object is to the Kerr black hole radius \cite{Wald1984}, $r_+$, such that $r=r_+(1+\epsilon)$ is the object's radius \cite{Cardoso2017}. For spherically symmetric objects $r_+=r_s$, where $r_s$ is the Schwarzschild radius.}, $\epsilon$, any measurable departures from the black hole geometry become inaccessible to a distant observer within a finite observation window\footnote{We consider infinite-time observations unphysical.}. As in Popper's black swan analogy \cite{Popper1959}, a single observation may rule out a specific model of an exotic compact object, yet \textit{no observational data is sufficient to confirm the existence of black holes}.

In this essay, we argue that the prevailing view that black holes have been definitely found in various astrophysical systems is epistemologically problematic. While we possess compelling black hole candidates in quasars, binary systems, and galactic nuclei, these remain exactly that: candidates. Observations may provide evidence consistent with black hole behavior, but they do not constitute confirmation of their existence. Certainly, black holes play an important role in several theoretical issues, and probing their properties is essential. Nonetheless, attention to what the data can truly reveal is critical to avoid an illusion of certainty and maintain scientific rigor. Critically, this issue is distinct from the question of whether black holes, as defined by general relativity, exist in our universe; it concerns the fundamental constraints general relativity places on what can be observationally confirmed.

\subsubsection*{Gravitational waves and ringdown}

Even before the detection of gravitational waves, the literature often suggested that such observations would provide definitive proof of black hole existence. For example, \cite[\S~1]{Berti2009} states:
\begin{quote}
    ``While electromagnetic observations are already providing us with strong evidence of the astrophysical reality of black holes, gravitational wave observations will \textit{incontrovertibly} show if these compact objects are indeed rotating (Kerr) black holes...'' [Italics added]
\end{quote}
Indeed, the first paper on gravitational wave detection adamantly asserted \cite[\S~VI]{Abbott2016}:
\begin{quote}
    ``GW150914 demonstrates the existence of stellar-mass black holes more massive than $\simeq 25\;M_{\odot}$, and establishes that binary black holes can form in nature and merge within a Hubble time.''
\end{quote}
Statements of this form are repeated for other events, e.g., GW170104 \cite[\S~I]{Abbott2018}:
\begin{quote}
    ``Detailed analyses demonstrate that GW170104 arrived at Hanford $\sim$ 3 ms before Livingston, and originated from the coalescence of two stellar-mass black holes...''
\end{quote}
The use of the word ``demonstrate'' in these contexts is epistemologically problematic; these compact coalescence events can, at most, serve as a profound investigation of merging binaries, rather than provide observational confirmation of black holes.

The primary argument supporting the interpretation that the result of compact coalescence is a black hole follows from the ringdown signal, i.e., the signature gravitational-wave behavior characterized by damped sinusoids arising from the dynamics of the post-merger remnant (see, e.g., \cite{Berti2009, Berti2025} for reviews). Namely, one finds that the gravitational-wave signal coincides with the quasinormal modes of a Kerr black hole, i.e., the characteristic oscillations of spacetime outside a perturbed Kerr black hole. Physically, this corresponds to the black hole reaching a stationary configuration.

A notorious difference between the quasinormal modes of a black hole and an extremely compact object is that, due to the presence of a surface on the latter, one expects echoes in the ringdown waveform. Naturally, one is interested in quantifying how subtle such a difference is. Notably, one finds that all black hole quasinormal data can be mimicked by an ultracompact object with an arbitrarily small closeness parameter \cite{Cardoso2016, Cardoso2017}. For a distant observer restricted to a finite observation window, the measurable differences between a black hole and a sufficiently compact horizonless model become inaccessible. Even if an observer could probe the near-horizon regime, the late-time modes of such objects typically carry insufficient energy to be distinguishable from the noise floor.

Consequently, while the ringdown waveform is remarkably consistent with the Kerr hypothesis, it reinforces the theory's predictive power rather than confirming the object's identity. Any compact object possessing a light ring will exhibit a similar ringdown stage; thus, the data provide evidence for the existence of light rings, not necessarily event horizons. Although light rings have been argued to be proxy evidence for black holes based on nonlinear stability arguments for ultracompact objects \cite{Cardoso2014}, this does not alter the fundamental conclusion. Even if a black hole appears to be the only viable solution to simultaneously explain all current data, say, by considering constraints on the equation of state of matter at high densities, consistency with the theory should not be taken as confirmation of its existence.  

Just as the 1919 Sobral eclipse expedition provided a successful test of light deflection without proving that spacetime is curved, modern ringdown observations are a triumphant test of general relativity that do not confirm the existence of black holes. Ultimately, these observations allow us to establish increasingly stringent upper bounds on $\epsilon$ and to rule out specific exotic models, but it is crucial to understand that they will never be sufficient to confirm the existence of black holes.

\subsubsection*{Shadows and Hawking radiation}

Electromagnetic observations have historically been at the vanguard of black hole research, from the identification of quasars to the recent breakthroughs of the Event Horizon Telescope collaboration. Notably, the interpretative stance adopted in the analysis of recent results from this collaboration, which show a dark, massive, and compact object, is a more cautious epistemological posture than that found in early gravitational-wave literature, acknowledging the inherent limits of their data \cite[\S~8]{EHT2019}:
\begin{quote}
``A number of elements reinforce the robustness of our image and the conclusion that it is consistent with the shadow of a black hole as predicted by GR [general relativity]. [...] At the same time, it is more difficult to rule out alternatives to black holes in GR, because a shadow can be produced by any compact object with a spacetime characterized by unstable circular photon orbits.'' 
\end{quote}
Despite several ``gastrophysical'' complexities, such as accretion flow and emission physics \cite{Gralla2021}, it is important to note that, at best, these provide evidence for black hole candidates rather than confirmation of black hole existence. This is reflected in the careful use of the term ``consistent'' and acknowledging candidates that could also explain the data, in contrast to stronger claims, such as ``demonstrate'', often found in the gravitational-wave literature.

Indeed, while the absence of a hard surface is expected to be a defining feature of black holes, the observed shadow is not unique to them. As quoted above, any sufficiently compact object with unstable circular photon orbits can produce a similar shadow. Namely, images such as those reported in \cite{EHT2019} do not constitute direct photographs of black holes, despite common interpretations; rather, they depict black hole candidates, contingent on astrophysical modeling assumptions. Hence, the celebrated ``first photo of a black hole'' is more accurately described as a high-resolution observation of a region of extreme gravitational lensing around a compact object. As with gravitational-wave observations, electromagnetic data can rule out specific models but cannot confirm the existence of black holes.

A related misconception concerns the potential detection of Hawking radiation (see, e.g., \cite[Ch.~4]{Parker2009} for an extensive pedagogical approach to the Hawking effect), an approximately thermal spectrum of particles, such that a collapsing body or ultracompact object effectively behaves as a gray body. Despite its theoretical elegance and foundational importance, the universality of the Hawking effect remains an open question (see, e.g., \cite{Unruh2005}), with unresolved issues such as the trans-Planckian problem, adiabaticity conditions, and assumptions underlying evaporation. More generally, Hawking-like radiation can be found in any case where one has an approximately exponential relation between the affine parameters on the null generators of past and future null infinity \cite{Harada2019, Barceló2011}. As a result, collapsing objects that do not form black holes and compact objects can produce similar particle spectra.

As with gravitational waves and shadows, the detection of Hawking-like radiation would merely set a lower bound on the compactness of the source. While such an observation would be a monumental triumph for semiclassical gravity, it would still fail to determine if the source is a black hole or a compact object. Moreover, these observations depend on additional, often unverified assumptions about the astrophysical environment. In this sense, no finite-time observation of radiation from a black hole candidate can determine whether it truly is a black hole.

\subsubsection*{Conclusion}

Taken together, one should exercise caution when interpreting gravitational-wave and electromagnetic data, particularly when using such observations to rule out ultracompact object models in favor of black holes as the only logical possibility. A more grounded approach is evident in some recent literature, such as the conclusions of the recent LIGO-Virgo-KAGRA collaboration \cite[Conclusion]{Abac2025}:
\begin{quote}
    ``Data from LIGO Hanford and LIGO Livingston are \textit{consistent} with multiple quasinormal modes of a remnant Kerr black hole and with Hawking’s area law. [...] Our results \textit{suggest} that astrophysical black holes are indeed extremely simple objects that follow general relativity and the Kerr description.'' [Italics added]
\end{quote}
The fact that one has overwhelming evidence that the observations are consistent with Kerr black holes in the framework of general relativity is an astonishing new test of the theory\footnote{Evidently, one needs to also consider unproven assumptions that are pertinent to obtain and interpret data.}. However, we must stress that these observations do not constitute definitive proof of the existence of black holes. While the distinction between evidence and confirmation is foundational to the philosophy of science, contemporary scientific reports occasionally succumb to linguistic abuse, whether due to misconceptions or for simplicity. All we can confidently affirm is that we are investigating black hole candidates, and the current data indicate that general relativity remains our most successful framework for describing them.

Future increases in the sensitivity of interferometers and telescopes will undoubtedly usher in an era of precision black hole astrophysics, allowing us to probe energy scales and regions closer to the purported surfaces or horizons of these candidates. Yet, regardless of the precision achieved, these observations will never provide conclusive proof that black holes exist.

Ultimately, a fundamental gap persists between theoretical elegance and observational reality. Beyond the current limitations of our technology, black holes face an even more intrinsic constraint imposed by general relativity itself. Let this essay serve as a call for scientific precision: theoretical consistency must never be advocated as empirical confirmation. Understanding the inherent boundaries of what a theory allows us to confirm is not a sign of doubt, but a requirement of rigor. We have not proven the existence of black holes; rather, we have accumulated compelling evidence for black hole candidates. Adopting a more disciplined language is essential to distinguish between the success of a mathematical model and the verified physical reality of black holes.

%%%%%%%%%%%%%%%%%%%%%%%%%%%%%%%%%%%%%%%%%%%%%%%%%%%%%%%%%%%%%%%%%%%%%%%%%%%%%%%%%%%%%%%%%%%%%%%%%%%%%%%%%%%%%%%%%%%%%%%%%%%%%%%%%%%
\subsubsection*{Acknowledgments}
The author thanks Daniel A. T. Vanzella for comments on a draft of this manuscript, extensive support, and stimulating discussions. This study was financed in part by the Coordenação de Aperfeiçoamento de Pessoal de Nível Superior – Brazil (CAPES) – Finance Code 001, and by the Sao Paulo Research Foundation (FAPESP), Brazil. Process Number 2024/19057-7.

\end{document}